\documentclass[aps,prc,twocolumn,showpacs,superscriptaddress]{revtex4-1}
\usepackage{natbib}
\usepackage{graphicx}
\usepackage{color}
\begin{document}
\title{Dipole response in $^{208}$Pb within  a self-consistent multiphonon approach}
\vspace{0.5cm}
\author{F. Knapp}
\affiliation{Faculty of Mathematics and Physics,  Charles University, Prague, Czech Republic  }
\author{N. Lo Iudice}
\affiliation{Dipartimento di Fisica, 
Universit$\grave{a}$ di Napoli Federico II, Italy}  
\affiliation{INFN Sezione di Napoli,  Italy}
\author{P. Vesel\'{y}}
\affiliation{Institute of Nuclear Physics,
Czech Academy of Sciences, Husinec - \v Re\v z, Czech Republic} 
\author{F. Andreozzi}
\affiliation{Dipartimento di Fisica, 
Universit$\grave{a}$ di Napoli Federico II,  Italy} 
\affiliation{INFN Sezione di Napoli, Italy} 
\author{G. De Gregorio}
\affiliation{Dipartimento di Fisica, 
Universit$\grave{a}$ di Napoli Federico II, Italy} 
\affiliation{INFN Sezione di Napoli, Italy}
\author{A. Porrino}
\affiliation{Dipartimento di Fisica, 
Universit$\grave{a}$ di Napoli Federico II,  Italy} 
\affiliation{INFN Sezione di Napoli,  Italy}
\date{\today}
\begin{abstract}
\begin{description}
\item[Background] The electric dipole strength detected around the particle threshold and commonly associated to the pygmy dipole resonance offers a unique information on  neutron skin and symmetry energy, and is of astrophysical interest. The nature   of such a   resonance is  still under debate.
\item[Purpose]  We intend to describe the giant and pygmy resonances in $^{208}$Pb by enhancing their fragmentation with respect to the random-phase approximation. 
\item[Method] We adopt the equation of motion phonon method to perform a fully self-consistent calculation in a space spanned by   one-phonon and two-phonon basis states using an optimized chiral two-body potential. A phenomenological density dependent term, derived from a contact three-body force, is added in order to get single-particle spectra more realistic than the ones obtained by using the chiral potential only.
The calculation takes into full account the Pauli principle and is free of spurious center of mass admixtures.
\item[Results] We obtain a fair description   of the giant resonance and obtain a dense  low-lying  spectrum in qualitative agreement with the experimental data. The transition densities as well as the phonon and particle-hole composition of the most strongly excited states support the pygmy nature of the low-lying resonance. Finally, we obtain realistic values for the dipole  polarizability and  the neutron skin radius. 
\item[Conclusions] The results emphasize the role of  the two-phonon states in enhancing the fragmentation of the strength in the giant resonance region and at low energy, consistently with experiments.   For a more detailed agreement with the data, the calculation suggests the inclusion of the three-phonon states as well as a fine tuning of the single-particle spectrum to be obtained by a refinement of the nuclear potential. 
\end{description}
\end{abstract}  
\maketitle
   
\section{Introduction} 

The electric dipole response in neutron rich nuclei was extensively investigated in recent years with special attention at
the low-energy  transitions around the neutron threshold associated  to the so called pygmy dipole resonance (PDR), a soft collective mode promoted by a translational oscillation of the neutron excess  against a N=Z  core \cite{Mohan71}.
So much attention is motivated by  the connection of the mode with the thickness of the neutron skin  and the symmetry energy 
and its relevance  to neutron star and other astrophysical phenomena.

These transitions were detected in several experiments on stable and unstable nuclei. An exhaustive list of references can be found in Ref.\cite{Savran13}.
Radioactive beam experiments have extracted an appreciable dipole strength just above the neutron decay threshold in unstable nuclei, like neutron rich oxygen \cite{Leiste} or tin isotopes around $^{132}$Sn \cite{Adri05}.
More detailed data were obtained for stable nuclei by  combining     $(\gamma, \gamma')$   
\cite{Gov98,Ryez02,Zilg02,Enders03,Hart04,Volz06,Ozel07,Schweng07,Isaak11,Savran11} with $(\alpha,\alpha' \gamma)$ measurements \cite{Savran06,Endres10,Endres12,Derya14} which have  detected  rich low lying  spectra of weakly excited discrete levels below   the  neutron threshold   over different mass regions.

Even more complete information was provided by a proton scattering experiment on $^{208}$Pb  which has  extracted   the   electric dipole spectrum  below and above the neutron threshold \cite{Tamii11,Polto,Polto14}.

The properties of the mode and its relevance to   neutron skin  and symmetry energy were investigated in a considerable number of theoretical approaches. We refer the reader to the reviews \cite{Savran13,Paar07,Paar10}.      Many calculations were carried out  in Hartree-Fock (HF) plus random-phase approximation (RPA)  \cite{Peru05,Piek06,Carbo10,Inak11,Lanza11,Piek11,Yuk12,Vret12} or, for open shell nuclei, Hartree-Fock-Bogoliubov (HFB) plus quasiparticle RPA (QRPA) \cite{Klimk07,Tera05,Tera06,Yosh08,Li08,Eba10,Marti11,Pap14}. 
Some of them  used Skyrme \cite{Tera05,Tera06,Carbo10} or Gogny forces \cite{Peru05,Marti11,Pap14}, others were carried out in  relativistic RPA (RRPA) using density functionals derived from meson-nucleon Lagrangians treated in mean field approximation 
\cite{Piek06,Piek11,Vret12}.

The fragmentation of the mode was faced in several  extensions,  like QRPA plus phonon coupling \cite{Sarchi04},  second RPA \cite{Gamba11}, the quasiparticle-phonon model  (QPM)
\cite{Ryez02,Tsoneva08,Savran11}, and the relativistic quasiparticle time-blocking approximation (RTBA) \cite{Litv0,Litv1,Litv2}.
   
We have proposed an equation of motion phonon method (EMPM)  \cite{AndLo,AndLo1,Bianco} which constructs and solves iteratively a set of equations of motion to generate   a multiphonon basis built of phonons obtained in Tamm-Dancoff approximation (TDA). Such a  basis simplifies greatly the structure of the Hamiltonian matrix and makes feasible its diagonalization in large configuration and phonon spaces.  

The formalism treats one-phonon as well as  multiphonon states on the same footing and 
takes the Pauli principle into full account.  Moreover, it holds for any realistic Hamiltonian of general type. 

It was already adopted to study the dipole response \cite{Bianco12,knap14}. In its most recent application, we performed a selfconsistent  calculation for the doubly magic $^{132}$Sn in a space spanned by one-phonon and two-phonon basis states \cite{knap14}.

We started with generating a HF basis using a   nucleon-nucleon (NN) chiral potential $V_\chi =NNLO_{opt}$ optimized so as to minimize the effects of the  three-body forces \cite{Ekstr13}.  A  review of the derivation and properties of chiral potentials can be found in Ref. \cite{Machl11}.

The HF spectrum so obtained resulted to be more compressed than the one obtained by using other realistic potentials \cite{HergPap11,Bianco14} but not sufficiently to approach closely the empirical single particle energies.  
This discrepancy was ascribed to the too attractive character of $V_\chi$ which overbinds Ca isotopes 
\cite{Ekstr13}, hinting that a, seemingly repulsive, contribution from the residual chiral three-body force is in order. 

Since such a term is not at our disposal, we added a phenomenological, density dependent, potential $V_\rho$ derived from a  contact  three-body interaction  \cite{Waroq} which improves the description of the  bulk properties in closed shell nuclei \cite{Gunter}.  When added to $V_\chi$, as to other NN potentials \cite{HergPap11,Bianco14}, it induces a further compression of the HF spectrum \cite{knap14}, though not enough to fill completely the gap with the empirical energies.  
As illustrated in Ref. \cite{knap14}, in fact, serious discrepancies between theory and experiments remain.   

The inclusion of $V_\rho$ enabled us to reproduce approximately the main peak of the GDR by a proper choice of the strength of $V_\rho$, the only parameter at our disposal.

The dipole spectra  resulted to be in rough agreement with the available data and  proved  the crucial role of the two-phonon basis states. The phonon coupling, in fact, has induced a strong fragmentation of the GDR and produced   a large number of  levels  around the neutron decay threshold.  
 
Here, we use the same modified  potential to perform an analogous selfconsistent EMPM calculation for $^{208}$Pb.
Due to the wealth of accurate data made available by the  $(p,p')$   \cite{Tamii11,Polto,Polto14},    $(\gamma,\gamma')$ and $(n,\gamma)$ \cite{Ryez02,Enders03,Shizu08,Schweng10} measurements combined with  the photo-absorption experiments \cite{Veyss70,Schel88}, this nucleus is a unique laboratory which allows to explore the detailed properties of  the dipole response and, in particular, the low-lying soft mode.  Its neutron skin was also expressly studied experimentally \cite{Tamii11,Abrah12,Tarbert14} and theoretically \cite{ReiNa10,Roca12,Piek12}. The nucleus, therefore, represents a testing ground for the  theoretical models.  
 
The EMPM was already adopted to study the dipole response in this nucleus \cite{Bianco12}.   In that paper, however, we used a modified oscillator single particle basis and  a Brueckner G-matrix \cite{Jens,Enge} derived from the CD-Bonn potential \cite{Mcl}. Moreover, the calculation was carried out in   more restricted configuration and phonon spaces.
This new calculation is also more satisfactory on theoretical ground. Apart from the coupling constant of the corrective density dependent term, we have no free parameters. Single-particle as well as  all other quantities are ultimately determined by the two-body potential.

\section{Brief description of the method}
\label{Eofm}

The Hamiltonian  we consider has the standard form
\begin{eqnarray}
H= H_0 + V.
\end{eqnarray}
In the $j-j$ coupled scheme, the one-body and two-body pieces assume the second quantized expressions
\begin{eqnarray}
H_0  =   \sum_{r} [r]^{1/2} \epsilon_r \Bigl(a^\dagger_r \times b_r\Bigr)^0,\\
V =  
- \frac{1}{4} \sum_{rsqt}^{ \Gamma} [\Gamma]^{1/2} V^\Gamma_{rsqt}\Bigl[ \Bigl(a^\dagger_r \times a^\dagger_s\Bigr)^\Gamma \times \Bigl(b_q \times b_t\Bigr)^\Gamma \Bigr]^0,\label{V}
\end{eqnarray}
where
\begin{eqnarray}
V^\Gamma_{rsqt} &=& < (q\times t)^\Gamma | V |(r \times s)^\Gamma> \nonumber\\
&&- (-)^{r+s- \Gamma} < (q\times t)^\Gamma | V |(s \times r)^\Gamma>.
\end{eqnarray}
Following French notation \cite{French}, we  have put $b_r = (-)^{j_r + m_r}a_{j_r - m_r}$,   $[\Gamma]= 2 \Gamma +1 = (2 J_\Gamma +1)$, and
\begin{equation}
\mid (r \times s)^\Gamma \rangle =  \sum_{m_r m_s} <j_r m_r j_s m_s|J_\Gamma M_\Gamma> \mid c_r j_r m_r c_s j_s m_s \rangle,
\end{equation} 
where $c_r$ denotes all the additional quantum numbers.
  
It is useful to write the   two-body potential (\ref{V}) in the recoupled form
\begin{eqnarray}
V=   \frac{1}{4}\sum_{rs qt  \sigma} [\sigma]^{1/2} F^\sigma_{rsqt} \Bigl[ \Bigl(a^\dagger_r \times b_s\Bigr)^\sigma \times \Bigl(a^\dagger_q \times b_t \Bigr)^\sigma \Bigr]^0, 
\end{eqnarray}
where
\begin{eqnarray}
F^\sigma_{rsqt}  =    \sum_{\Gamma}  [\Gamma]  (-)^{r+t - \sigma -\Gamma} W(rsqt;\sigma \Gamma)
  V^\Gamma_{rqst} 
\end{eqnarray}
and $W(rsqt;\sigma \Gamma)$ are Racah coefficients. 

The primary goal of the method is to generate an orthonormal  basis of  $n$-phonon states built of TDA phonons. 
Let us assume that the $(n-1)$-phonon basis states $ \mid n-1, \alpha>$ are known. We must, then, 
determine the $n$-phonon basis states of the form 
\begin{equation}
|n;\beta> = \sum_{ \lambda \alpha} 
C_{\lambda \alpha }^{\beta } \, \Bigl\{O^\dagger_\lambda \times \mid n-1, \alpha> \Bigr\}^\beta,  
\label{nstatec}
\end{equation}
where 
\begin{eqnarray}
 O^\dagger_{\lambda}    =  \sum_{ph} c^\lambda_{ph}  ( a^\dagger_p \times b_h)^\lambda
\label{Olam} 
\end{eqnarray}
is the TDA particle-hole  (p-h) phonon operator.
We start with  the equations of motion
\begin{eqnarray}
&&<n, \beta \mid \Bigl\{\Bigl[H, O^\dagger_\lambda \Bigr] \times \mid n-1, \alpha >\Bigr\}^\beta = \nonumber\\
&& \Bigl(E_\beta - E_\alpha \Bigr) 
 <n, \beta \mid \Bigl\{O^\dagger_\lambda \times \mid n-1, \alpha> \Bigr\}^\beta.\nonumber\\
\end{eqnarray}
Upon applying the Wigner-Eckart theorem, we obtain the equivalent equations
\begin{eqnarray}
 &&<n, \beta \parallel [H, O^\dagger_\lambda]   \parallel n-1, \alpha >  = \nonumber\\
&&\Bigl(E_\beta - E_\alpha \Bigr)  
<n, \beta \parallel O^\dagger_\lambda  \parallel n-1, \alpha>.
\end{eqnarray}
As explained in Ref. \cite{Bianco}, we  expand  the commutator and    invert   Eq. (\ref{Olam}) in order to express the p-h  operators,  present in the expanded commutator,  in terms of the phonon operators $O^\dagger_{\lambda}$. We obtain   
\begin{eqnarray}
 \sum_{\lambda' \gamma}  {\cal A}^{\beta} (\lambda \alpha, \lambda' \gamma) X^\beta_ {\lambda' \gamma} 
= E_\beta   X^\beta_ {\lambda \alpha},
\label{AX}  
\end{eqnarray}
where $X$ defines the amplitude
\begin{eqnarray}
X^\beta_ {\lambda \alpha}=  <n, \beta \parallel O^\dagger_\lambda  \parallel n-1, \alpha> \label{Xc}
\end{eqnarray}
and ${\mathcal A}$ is a matrix of the simple structure
\begin{eqnarray}
{\mathcal A}^{\beta} (\lambda \alpha, \lambda' \gamma)  &= & 
  (E_\lambda + E_\alpha) \delta_{\lambda \lambda'} \delta_{\alpha \gamma}  \nonumber\\ 
 && + \sum_{  \sigma  }     W(\beta \lambda' \alpha \sigma; \gamma \lambda)  {\cal V}^\sigma_{\lambda \alpha, \lambda' \gamma}.   
\label{Anphc}
\end{eqnarray}
Here, the phonon-phonon potential is given by
\begin{eqnarray}
{\cal V}^\sigma_{\lambda \alpha, \lambda' \gamma}    =  \sum_{  rs}    
{\cal V}^\sigma_{\lambda \lambda'}(rs) \rho^{(n)}_{\alpha \gamma} ([r\times s]^\sigma), \label{Vphon}
 \end{eqnarray}
where the labels $(r s)$ run   over particle $(rs = p p')$ and hole $(rs = h h')$ states.
In the above equation, we have introduced  the $n$-phonon density matrix
\begin{eqnarray}
\rho^{(n)}_{\alpha \alpha'} \bigl([r \times s]^\sigma \bigr)  =
 \langle n; \alpha'    \parallel \Big[a^\dagger_ {r } \times b_{s} \Bigr]^{\sigma}  \parallel n; \alpha  \rangle  
\label{rhonph}
\end{eqnarray}
and the potential 
\begin{eqnarray}
{\cal V}^\sigma_{\lambda \lambda'} (rs)    =     
\sum_{tq} \rho_{\lambda \lambda'} \bigl([q \times t]^\sigma \bigr) 
   F^\sigma_{qt rs},  \label{v1ph }    
  \end{eqnarray}
where $\rho_{\lambda \lambda' }$ is the TDA density matrix given by 
\begin{eqnarray}
\rho_{\lambda \lambda' } ([r\times s]^\sigma)  =  <\lambda' \parallel  \Bigl( a^\dagger_r \times b_s \Bigr)^\sigma \parallel \lambda>  \nonumber\\
=[\lambda \lambda' \sigma]^{1/2} \sum_{t  }  c^\lambda_{ts}  c^{\lambda'}_{tr} 
W(\lambda' t \sigma s; r \lambda). \label{rho1ph} 
\end{eqnarray}
Here $t$ runs over particle  ($t=p$) or hole ($t=h$) states according that $(rs)=(h h')$ or $(rs)=(pp')$, respectively.

The formal analogy between the structure  of the phonon matrix ${\cal A}^\beta (\lambda \alpha,  \lambda' \alpha')$   and the form of the TDA matrix $A^\lambda (ph;p'h')$  was pointed out in Ref. \cite{Bianco}. The first is deduced from  the second by replacing the p-h energies with the sum of   phonon energies and the p-h interaction with a phonon-phonon interaction. 

Eq. (\ref{AX}) is not an eigenvalue equation yet. We have first to expand the amplitudes $X$ (Eq. \ref{Xc}) in terms of the expansion coefficients $C_{\lambda \alpha }^{\beta }$ of the states $|n;\beta>$ (Eq. (\ref{nstatec})) obtaining 
\begin{eqnarray}
  X_{\lambda \alpha}^{\beta} = 
   \sum_{\lambda' \alpha' } {\cal D}^{\beta}  (\lambda \alpha,\lambda' \alpha' )C_{\lambda' \alpha' }^{\beta }, 
  \label{XC}
\end{eqnarray}
where
\begin{eqnarray}
  {\cal D}^{\beta}  (\alpha \lambda ; \alpha' \lambda' ) =
\nonumber\\ =   
 \Bigl[ < n-1, \alpha \mid \times O_{\lambda} \Bigr]_{\beta } \Bigl[ O^\dagger_{\lambda'} \times \mid n-1, \alpha'>\Bigr]_{\beta }
\end{eqnarray}
is the metric matrix which reintroduces the exchange terms among different phonons and, therefore, re-establishes   the Pauli principle. 

Upon insertion of the expansion (\ref{XC}) into Eq. (\ref{AX}), we get the generalized eigenvalue equation
\begin{eqnarray}
   \sum_{\lambda' \alpha' }  {\cal H}^{\beta} (\lambda \alpha, \lambda' \alpha')  C^\beta_ {\lambda' \alpha'}  = 
   \nonumber \\
    \sum_{\lambda' \alpha' } (A  {\cal D})^{\beta} (\lambda \alpha, \lambda' \alpha')  C^\beta_ {\lambda' \alpha'}=
 E_\beta \sum_{\lambda' \alpha'  }    {\cal D}^{\beta} (\lambda \alpha, \lambda' \alpha')C^\beta_ {\lambda' \alpha'}.  
 \label{Eig}
\end{eqnarray}
This is effectively the representation of the eigenvalue equation 
\begin{equation}
  {\cal H} C =  ( {\cal A}   \, {\cal D})    C   = E  C   
\label{Eigmat}
\end{equation}
in the overcomplete basis $\Bigl\{O^\dagger_\lambda \times \mid n-1, \alpha> \Bigr\}^\beta$.

We eliminate such a redundancy by following the procedure outlined in Ref. \cite{AndLo,AndLo1}, based on the Cholesky decomposition method.
This method  selects a basis of linear independent states 
$\Bigl\{O_{\lambda}^\dagger \times |n-1;\,\alpha>\Bigr\}^\beta$ spanning the physical subspace 
of the correct dimensions $N_n < N_{r}$ and, thus, enables us to 
construct a $N_n \times N_n$ non singular matrix 
$ {\cal D}_{n}$. By left multiplication in the $N_n$-dimensional subspace
we get from Eq. (\ref{Eigmat}) 
\begin{equation}
\left[ {\cal D}_n^{-1}  {\cal H}  \right] C   = \left[ {\cal D}_n^{-1} ( {\cal A}   \, {\cal D})  \right] C =E  C.  
\label{Eigcho}
\end{equation}
This equation determines only the coefficients  $ C^{\beta}_{\lambda \alpha }  $ of the $N_n$-dimensional physical subspace.
The remaining redundant $N_{r}-N_n$ coefficients are undetermined and,  therefore, can  be safely put equal to zero. 
The eigenvalue problem within the $n$-phonon subspace is thereby solved exactly and yields a basis of orthonormal correlated $n$-phonon states of the form (\ref{nstatec}).

Since recursive formulas hold for all quantities entering  ${\cal A}$ and ${\cal D}$, it is possible to solve the eigenvalue equations    iteratively starting from the TDA phonons and, thereby, generate a set of orthonormal multiphonon states $\{|0>, |1, \lambda>,  \dots |n,\alpha> \dots \}$.

In such a basis, the Hamiltonian matrix is composed of a sequence of diagonal blocks, one for each $n$, mutually coupled by  off-diagonal terms $\langle n' \mid H \mid n \rangle$ which are non vanishing only for $n' = n \pm 1, n\pm2$ and are computed by means of recursive formulas.  A matrix of such a simple structure can be  easily diagonalized  yielding    
eigenfunctions of  the form
\begin{eqnarray}
\mid \Psi_{\nu} \rangle  =  \sum_{n \alpha} C^{(\nu)}_{\alpha} \mid n; \alpha \rangle .
\label{Psi}
\end{eqnarray}
These eigenfunctions are  used to compute the transition amplitudes. 
In the coupled scheme, the one-body operator has the form
\begin{eqnarray}
{\cal M} (\lambda)  = \frac{1}{[\lambda]^{1/2}}
\sum_{rs} <r \parallel  {\cal M} (\lambda) \parallel   s > \Bigl[a_r^\dagger \times b_s \Bigr]^{\lambda}. 
\end{eqnarray} 
The reduced transition amplitudes   are given by 
\begin{eqnarray}  
\langle \Psi_{f J_f } \parallel {\cal M} (\lambda) \parallel \Psi_{i J_i} \rangle =\nonumber\\ 
\sum_{(n_i \alpha) (n_f \beta)}  C^{(i)}_{n_i \alpha}
C^{(f)}_{ n_f \beta } <n_f, \beta J_f \parallel {\cal M} (\lambda) \parallel n_i, \alpha J_i>.
\label{Mlamampl}
\end{eqnarray}
The matrix elements of ${\cal M} (\lambda) $ between multiphonon states are  
\begin{eqnarray}
\label{Mnab}
<n_f;\beta J_f  \parallel  {\cal  M} (\lambda) \parallel  n_i; \alpha J_i   >  \nonumber\\
= [\lambda]^{-1/2} 
\Biggl[ \delta_{n_f (n_i+1)}  \sum_{x } {\cal  M}_\lambda (0 \rightarrow x)  X^{\beta}_{(x \lambda) \alpha}  
\nonumber\\
+ \delta_{n_f (n_i-1)}   (-)^{J_f   - J_i   } \sum_{x } {\cal  M}_\lambda (0 \rightarrow x)  X^{\alpha}_{(x \lambda) \beta }
\nonumber\\
 + \delta_{n_f n_i} {\cal  M}^{(n_i)}_{\alpha \beta}(\lambda)  
\Biggr].
\end{eqnarray}
where  
\begin{eqnarray}
\label{MlTDA}
{\cal  M}_\lambda (0 \rightarrow x) = < x \lambda  \parallel  {\cal  M} (\lambda) \parallel  0  >
=\nonumber\\
= \sum_{ph} c^{(x \lambda)}_{ph} <p \parallel {\cal  M} (\lambda)   \parallel h>
\end{eqnarray}
is the TDA transition amplitude and 
\begin{eqnarray}
\label{Mnn}
{\cal  M}^{(n_i)}_{\alpha\beta}(\lambda) =
\sum_{r s} <r \parallel {\cal  M} (\lambda) \parallel s>  \rho^{(n_i)}_{\alpha \beta}
\left(\left[r \times s \right]^{\lambda}\right)
\end{eqnarray}
the scattering term between states with the same number of phonons $n_i$.

\section{Implementation of the method} 
\label{calc}
\subsection{Hamiltonian}  
The Hamiltonian we used has the form
\begin{equation}
H= T_{int} + V.   
\end{equation}
Here 
\begin{equation} 
\label{Tint}
T_{int} =  \frac{1}{2m} \sum_i p_i^2  - T_{CM} 
\end{equation} 
is the intrinsic kinetic operator and 
\begin{equation}
V =   V_\chi + V_\rho
\end{equation}
is a two-body potential composed of two terms. The first is the    $NN$ optimized chiral potential $V_\chi$ = NNLO$_{opt}$  
determined in  Ref.  \cite{Ekstr13} by fixing the  coupling constants   at next-to-next leading order through a new optimization  method in the analysis of    the phase shifts, which minimizes the effects of the three-nucleon force. 

Though successful in reproducing   several bulk and spectroscopic properties of light and medium-light nuclei, $V_\chi$ overestimates the binding energy per nucleon of Ca isotopes by about 1 MeV and, most likely, overbinds even more the heavier nuclei.  
In order to offset its over attractive action we add the phenomenological density dependent term
\begin{equation}  
V_{\rho} = \sum_{i < j} v_\rho (ij),
\end{equation}
where
\begin{eqnarray}
v_\rho  = \frac{C_\rho}{6} (1 + P_\sigma) \rho(\frac{\vec{r}_1 + \vec{r}_2   }{2}) \delta(\vec{r}_1 - \vec{r}_2).
 \end{eqnarray}
This potential was derived  in Ref. \cite{Waroq} from a three-body contact interaction  
\begin{equation}
v_3 = C_\rho  \delta(\vec{r}_1 - \vec{r}_2)  \delta(\vec{r}_2 - \vec{r}_3)
\label{V-3body}
\end{equation}
which improved the description of bulk properties in closed shell nuclei
within a HF plus perturbation theory approach \cite{Gunter}.    

By adding $V_{\rho}$ to realistic potentials deduced from  nucleon-nucleon forces  \cite{HergPap11,Bianco14,knap14}, it was possible to improve partially the agreement of the  single particle  spectra with experiments and to approach the observed main peak of the GDR. The effects of $V_\rho$ will be further discussed in the next section.

\subsection{Spurious admixtures}
\label{spurious}
Dealing with low-energy dipole spectra, it is of great importance to obtain $1^-$ states free of spurious admixtures induced by the CM excitation.
We achieve this task for the TDA $1^-$ phonons by the method outlined in Ref. \cite{Bianco14} based on the Gramm-Schmidt orthogonalization method. The method yields states$\mid \varphi_r\rangle$ which are linear combinations of p-h states and  are orthogonal to the 
CM spurious state 
\begin{eqnarray}
\mid \varphi_0 \rangle =   \frac{1}{N_1} R_{\mu} \mid 0 \rangle, 
\end{eqnarray}
where $R_{\mu}$ is the CM coordinate and $N_1$ the normalization constant.
 
These basis states   are adopted to construct and diagonalize the Hamiltonian matrix. The resulting  eigenstates are  rigorously free of spurious admixtures   induced by the CM excitation.  They are then expressed  in terms of  the $\mid (p \times h^{-1})^{1^-} \rangle$  states to recover the standard TDA structure.
 
The EMPM states built of these CM spurious free TDA phonons are also spurious free. They are linear combinations of products of  phonons. The exchange terms between Fermions entering two different phonons of a product are taken into account through the metric matrix $D$ which leaves the phonon structure unchanged.

 \begin{figure}[ht]
\includegraphics[width = \columnwidth]{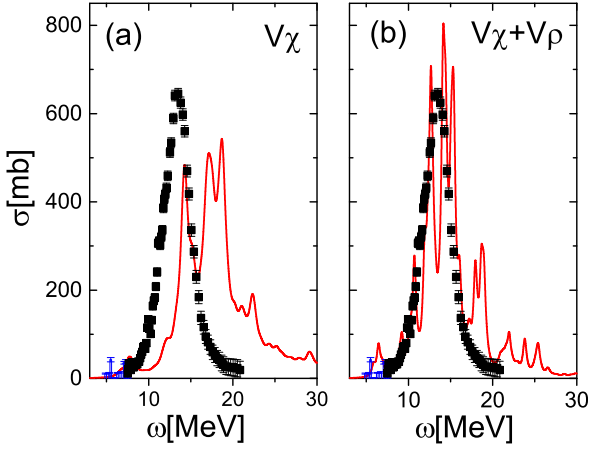}
\caption{ (Color online)  \label{fig1}  TDA  $E1$ cross section in    $^{208}$Pb  computed using     $V_{\chi}$ (a) only and   $V_{\chi} + V_\rho$ (b). The experimental data (black squares) are taken from \cite{Tamii11,Polto,Schweng10,Veyss70,Schel88}.   
} 
\end{figure}

\begin{figure}[ht]
\includegraphics[width=\columnwidth]{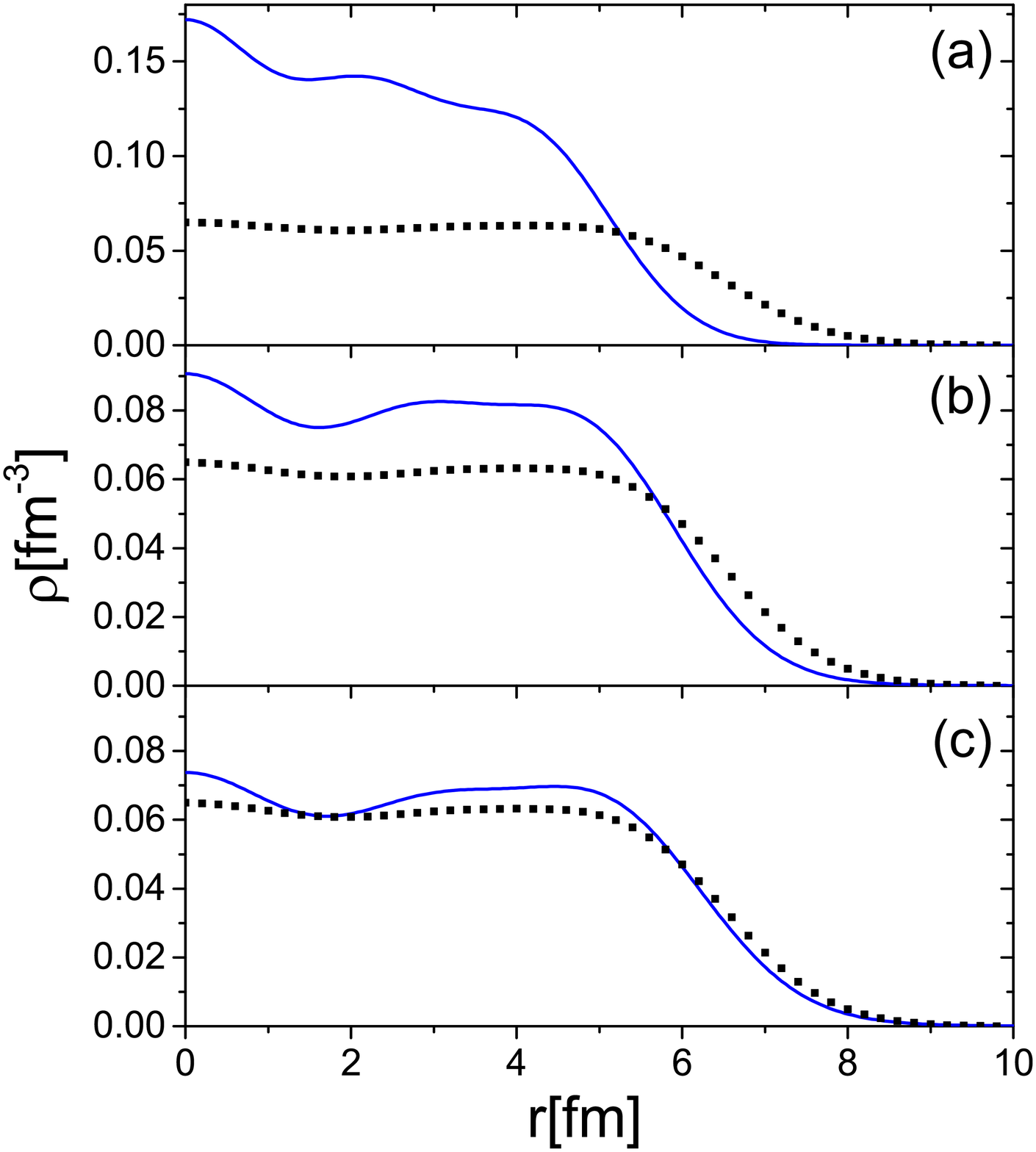}
\caption{(Color online)   \label{fig2} 
Comparison with experiments \cite{DeVries87} of the HF charge densities obtained by using   for the strength of $V_\rho$  the values (a) $C_\rho =0$, (b) $C_\rho =2000$ MeV fm$^6$, (c) $C_\rho =3000$ MeV fm$^6$ }  
\end{figure}

\section{Calculations and results}
\subsection{Dipole response} 
Our main goal is to compute   the $E1$ reduced strength
\begin{equation}
B_{\nu} (E \lambda=1 )=  
 \left| \langle  \nu  \parallel {\cal M}(E \lambda =1)   \parallel
 0^+ \rangle \right|^{2},   \label{BE}
\end{equation}
where    
\begin{eqnarray}
\label{Mla}
{\cal M} (E \lambda \mu) = \frac{e}{2}
\sum_{i=1}^A (1-\tau_3^i ) r^\lambda_i Y_{\lambda \mu}(\hat{r_i})
\label{Elamb}
\end{eqnarray}
is the electric multipole operator written as the sum of its isoscalar and isovector components, with $\tau_3 = 1$ for neutrons and $\tau_3 = -1$  for protons.

By inserting  the wave functions (\ref{Psi}) up to two phonons into the formula (\ref{Mlamampl})
and  making use of Eq. (\ref{Mnab}), we get for the $\lambda =1$ transitions from the ground to the $1^-$ states  
\begin{eqnarray}
\label{M0l2}
&&<\nu \parallel  {\cal M}(E \lambda) \parallel   0^+> = 
{\cal C}^{(0^+)}_{0}  \sum_{  \beta_1}{\cal C}^{(\nu)}_{\beta_1} {\cal  M}_\lambda (0^+ \rightarrow \beta_1 )  \nonumber\\
&& + (-)^\lambda  [\lambda]^{-1/2}  \sum_{x} {\cal  M}_\lambda (0^+ \rightarrow x   ) Y_{x \lambda} (0^+ \rightarrow \nu)
 \nonumber\\
&& +[\lambda]^{-1/2} \sum_{  \alpha_2  \beta_2  } {\cal C}^{(0^+)}_{\alpha_2} {\cal C}^{(\nu \lambda)}_{\beta_2} {\cal M}_{\alpha_2 \beta_2} (\lambda)  
\end{eqnarray}
where $\alpha_n = (n \alpha)$ and $\beta_n = (n \beta)$ label the $n$-phonon components $\mid n \alpha \rangle$ and $\mid n \beta \rangle$ of the ground $\Psi^{0^+}_g$ and final $1^-$ states $\Psi_\nu$, respectively. The quantity
${\cal  M}_\lambda (0^+ \rightarrow \beta_1 )$ is the TDA amplitude (\ref{MlTDA}) of the transition   to the one-phonon component $\mid \beta_1 \lambda \rangle$ of the final state $\Psi_\nu$ 
and 
\begin{eqnarray}
Y_{x \lambda} (0^+ \rightarrow \nu) =\sum_{\alpha_2 \beta_1}   {\cal C}^{(0^+)}_{\alpha_2} {\cal C}^{(\nu)}_{\beta_1 }  
X^{(\alpha_2)}_{(x \lambda) (\beta_1 \lambda)},   
\end{eqnarray} 
where $X^{(\alpha_2)}_{(x \lambda) (\beta_1 \lambda)} = \langle \alpha_2, 0^+ \parallel O^\dagger_{x \lambda} \parallel \beta_1 \lambda \rangle$.

The first term  is dominant. The second  may affect transitions involving $1^-$ and ground states having sizable one-phonon and two-phonon components, respectively, while the third may contribute to the transitions connecting states both having appreciable two-phonon components.

We will compute the amplitudes defined above, by using the EMPM correlated as well as the unperturbed HF ground state. In this latter case, the formula  becomes simply 
\begin{eqnarray}
\label{M01ph}
<\nu \parallel  {\cal M}(E \lambda) \parallel   0^+> = 
\sum_{  \beta_1}{\cal C}^{(\nu)}_{\beta_1} {\cal  M}_\lambda (0^+ \rightarrow \beta_1 ).  
\end{eqnarray}

The $B_{\nu} (E1 )$ strength is used to determine the dipole cross section
\begin{eqnarray}
\sigma  =  \int_{0}^{\infty } \sigma (\omega) d\omega = \frac{16 \pi^3}{9\hbar c} \int_{0}^{\infty } \omega {\cal S}(E1, \omega) d\omega 
 \label{sigma}
\end{eqnarray}
where ${\cal S}(E1, \omega)$ is the  $\lambda =1$  strength function   
\begin{eqnarray}
{\cal S}(E\lambda, \omega) = \sum_{\nu }B_{\nu}  ( E\lambda )\,\delta (\omega -\omega_{\nu })
\nonumber\\
\approx  
\sum_{\nu }B_{\nu}  ( E\lambda )\,\rho _{\Delta }(\omega -\omega_{\nu }).
\label{Strength}
\end{eqnarray}
Here  $\omega$ is the energy variable,  $\omega_\nu$ the energy of the transition of multipolarity $E\lambda$   from the ground to the $\nu_{th}$ excited state   of spin $J = \lambda$ and
\begin{equation}
\rho _{\Delta }(\omega - \omega_\nu)=\frac{\Delta}{2\pi } \frac{1}{(\omega - 
\omega_\nu)^{2}+(\frac{\Delta}{2})^{2}}  \label{weight}
\end{equation}
is  a Lorentzian  of width $\Delta$, which replaces the $\delta$ function as a
weight of  the reduced strength.

After integration, the cross section becomes
\begin{eqnarray}
\sigma  =   \frac{16 \pi^3}{9\hbar c} m_{1}, 
 \label{sigma1}
\end{eqnarray}
where 
\begin{eqnarray}
 m_{1} = \sum_\nu \omega_\nu B_\nu (E1)
\label{EWS}
\end{eqnarray}   
is the first moment. If one neglects momentum dependent and exchange terms in the Hamiltonian, $m_1$  fulfills the classical energy weighted Thomas-Reiche-Kuhn (TRK) sum rule
\begin{eqnarray}
 m_{1}= \frac{\hbar^2}{2 m } \frac{9}{4 \pi} \frac{N Z}{A} e^2 
 \label{TRK}
\end{eqnarray}
and the total cross section   assumes the value
\begin{eqnarray}
\sigma  =      (2 \pi)^2 \frac{\hbar^2}{2m} \frac{e^2}{\hbar c}  \frac{N Z}{A} 
= 60 \frac{N Z}{A} ({\rm MeV} {\rm mb}). 
 \label{sigmaTRK}
\end{eqnarray}
An observable sensitive to the low-energy spectrum is the dipole polarizability \cite{ReiNa10,Piek11} 
\begin{eqnarray}
 \alpha_D = \frac{8 \pi}{9} m_{-1},
 \label{alphad}
\end{eqnarray}
where
\begin{eqnarray}
 m_{-1} = \sum_n \omega^{-1}_\nu B_\nu (E1)
\label{m-1}
\end{eqnarray}
is the $E1$ inverse moment.
This quantity links the dipole polarizability   to the cross section $\sigma (\omega)$ according to the relation \cite{Tamii11}
\begin{eqnarray}
\alpha_D  = \frac{\hbar c}{2 \pi^2 e^2} \int_{0}^{\infty } \omega^{-2} \sigma (\omega) d\omega.  
 \label{alfaD}
\end{eqnarray}

\subsection{Hartree-Fock and TDA}

The HF basis is generated in a configuration space which includes   13 harmonic oscillator major shells, up to the principal quantum number $N_{max}=12$. 
This space is sufficient for reaching a good convergence of the single particle spectra below and around the Fermi surface. In going, for instance, from $N_{max} =9$ to $N_{max} =12$, step by step,  the energies change at most by $\sim 0.1$ MeV  at each step.  
More appreciable variations with $N_{max}$ are noticed in the spectrum far above the Fermi surface, a general feature of HF. They, however, induce some fluctuations only   in   the high energy  sector of the TDA strength distribution, which are wiped out by   the coupling with  the two-phonon states. The low-energy dipole spectrum reaches convergence very fast even at the TDA level. The rate of convergence for both HF and TDA is the same whether we add or not $V_\rho$ to  $V_{\chi}$.

We have generated the TDA phonons  $\mid \lambda \rangle$  in a space which encompasses three major shells above and only one major shell below the Fermi surface. We have checked that this is the minimal space needed to get solutions very close to the ones resulting from diagonalizing the Hamiltonian in the full HF space.

Though  $V_{\chi}$ yields considerably more compressed HF spectra \cite{knap14} compared to other potentials \cite{Bianco14,HergPap11}, the gap between major shells remains too large with respect to the empirical single particle levels. Due to this large gap, the TDA dipole cross section  gets peaked   $\sim 7$ MeV above the experimental peak (Figure \ref{fig1}). The two peaks almost overlap once we add  $V_{\rho}$ with a coupling constant  $C_\rho \sim 2000$ MeV $fm^6$. 

The use of $V= V_\chi +V_\rho$  improves considerably also the agreement between  the HF and  the measured nuclear charge density distribution  (Figure \ref{fig2}).  The sensitivity to $V_\rho$ is to be noticed. The empirical charge distribution is reproduced fairly well for $C_\rho \sim 3000$ MeV $fm^6$. It is, however,  more appropriate to use a smaller value since additional diffuseness of the Fermi surface is to be expected from the ground state correlations. We, thus, keep the value $C_\rho \sim 2000$ MeV $fm^6$ suggested by the dipole response. 

It must be stressed once again that, as shown in Ref. \cite{knap14}, the HF spectrum so obtained still differs significantly from the empirical one. The consequences of this discrepancy on the dipole strength distribution will be discussed later.

\subsection{EMPM}
The two-phonon  basis states  $\mid (n=2) \beta \rangle$ are generated in a truncated space spanned by the states $\mid (\lambda_1 \times \lambda_2)^{\beta} \rangle \equiv \Bigl\{O^\dagger_{\lambda_1} \times \mid \lambda_{2}> \Bigr\}^{\beta}$ composed of all TDA phonons, of both parities and of all multipolarities,  with dominant $1 \hbar \omega$ particle-hole components and, in addition,  all two-phonon states with  $E_{\lambda_1} + E_{\lambda_2} \alt 20$ MeV and $E_{\lambda_i} \alt 15 MeV$. 

This restricted set is used    to construct and diagonalize the Hamiltonian matrix. After the Cholesky treatment, we  obtain a truncated basis of $\sim$2.500 and $\sim$6.000 correlated orthonormal  states $\mid (n=2) \beta, 0^+ \rangle$ and $\mid (n=2) \beta, 1^- \rangle$, respectively.  

No significant changes in the spectrum are produced if  the number of TDA phonons is increased. We have raised the acceptance limit up to 30 MeV
 ($E_{\lambda_1} + E_{\lambda_2} \alt 30$ MeV)   thereby generating $\sim$9.000  and $\sim$22.000 $\mid (n=2) \beta, 0^+ \rangle$ and $\mid (n=2) \beta, 1^- \rangle$ states, respectively. We found that the low-lying energy peaks are shifted downward by only $\sim 0.2$ MeV and the strength distribution is little affected. 

The correlated $\mid (n=2) \beta\rangle$  states are added to the  unperturbed ground state plus the TDA one-phonon basis to diagonalize the residual Hamiltonian and determine   the ground and $1^-$ EMPM states.

Due to the two-phonon coupling, the ground state gets    depressed by  $\Delta E = 7.3$ MeV  with respect to the unperturbed HF state and becomes  strongly correlated. Its two-phonon components account for $\sim 24 \%$ of the wavefunction, in analogy with previous results obtained   for $^{16}$O \cite{Bianco}, the same $^{208}$Pb \cite{Bianco12} and $^{132}$Sn \cite{knap14}  and, also,  consistently with shell model calculations on  $^{208}$Pb \cite{Schweng07,Brown}. 

This ground state energy shift would spoil the description of the dipole response by pushing the strength at too high energy,  unless we add the three-phonon basis states to determine  the $1^-$ eigenstates. It is in fact known from EMPM calculations on $^{16}$O \cite{Bianco} that the three-phonon configurations couple strongly to the $1^-$ TDA phonons and, thereby, counterbalance the coupling between ground and two-phonon states by pushing the strength back to the experimental region.   
 
Including the three-phonon basis in a heavy nucleus like $^{208}$Pb is too time consuming, without drastic and uncontrollable truncations and approximations. Thus, we confine ourselves to a two-phonon space and, for consistency,  refer the energies to the unperturbed HF ground state. 
This assumption, implicit in all extensions of RPA, was already made  in  shell model calculations \cite{Schweng07,Brown} as well as in our previous EPMP calculations  \cite{Bianco12,knap14}.

\begin{figure}[ht]
\includegraphics[width=\columnwidth]{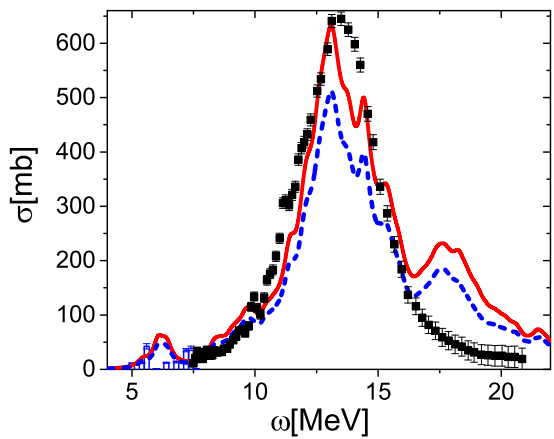}
\caption{ (Color online)  \label{fig3} Experimental \cite{Tamii11,Polto,Schweng10,Veyss70,Schel88} versus   EMPM   $E1$ cross sections in $^{208}$Pb. A Lorentzian    of width $\Delta = 0.5$ MeV was adopted. The EMPM cross section is computed using a HF (continuous red line) and a correlated (dashed blue line) ground states.
 } 
\end{figure}

\begin{figure}[ht]
\includegraphics[width = \columnwidth]{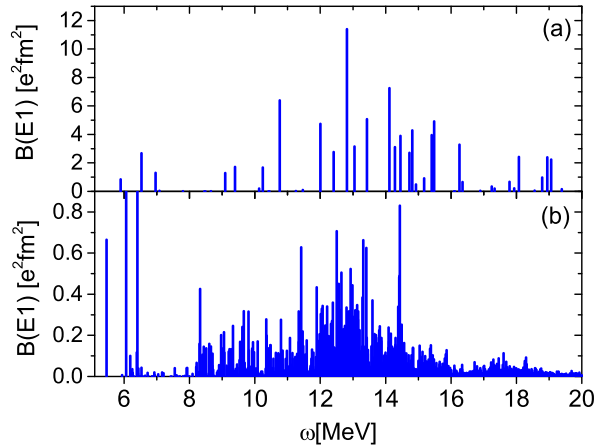}
\caption{ (Color online)  \label{fig4}   TDA (a) versus EMPM (b) $E1$ strength distributions   in    $^{208}$Pb. 
The different scales used for the two-plots are to be noticed.      
} 
\end{figure}

\begin{figure}[ht]
\includegraphics[width = \columnwidth]{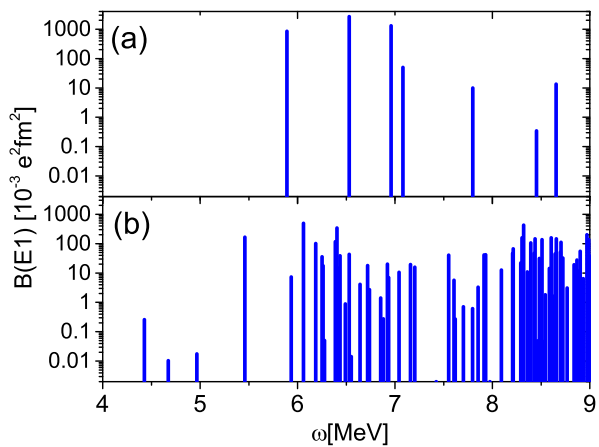}
\caption{(Color online)   \label{fig5}  TDA (a) versus EMPM (b) $E1$ low-lying spectra 
 plotted on an amplified (logarithmic) scale.
} 
\end{figure}

\begin{figure}[ht]
\includegraphics[width = \columnwidth]{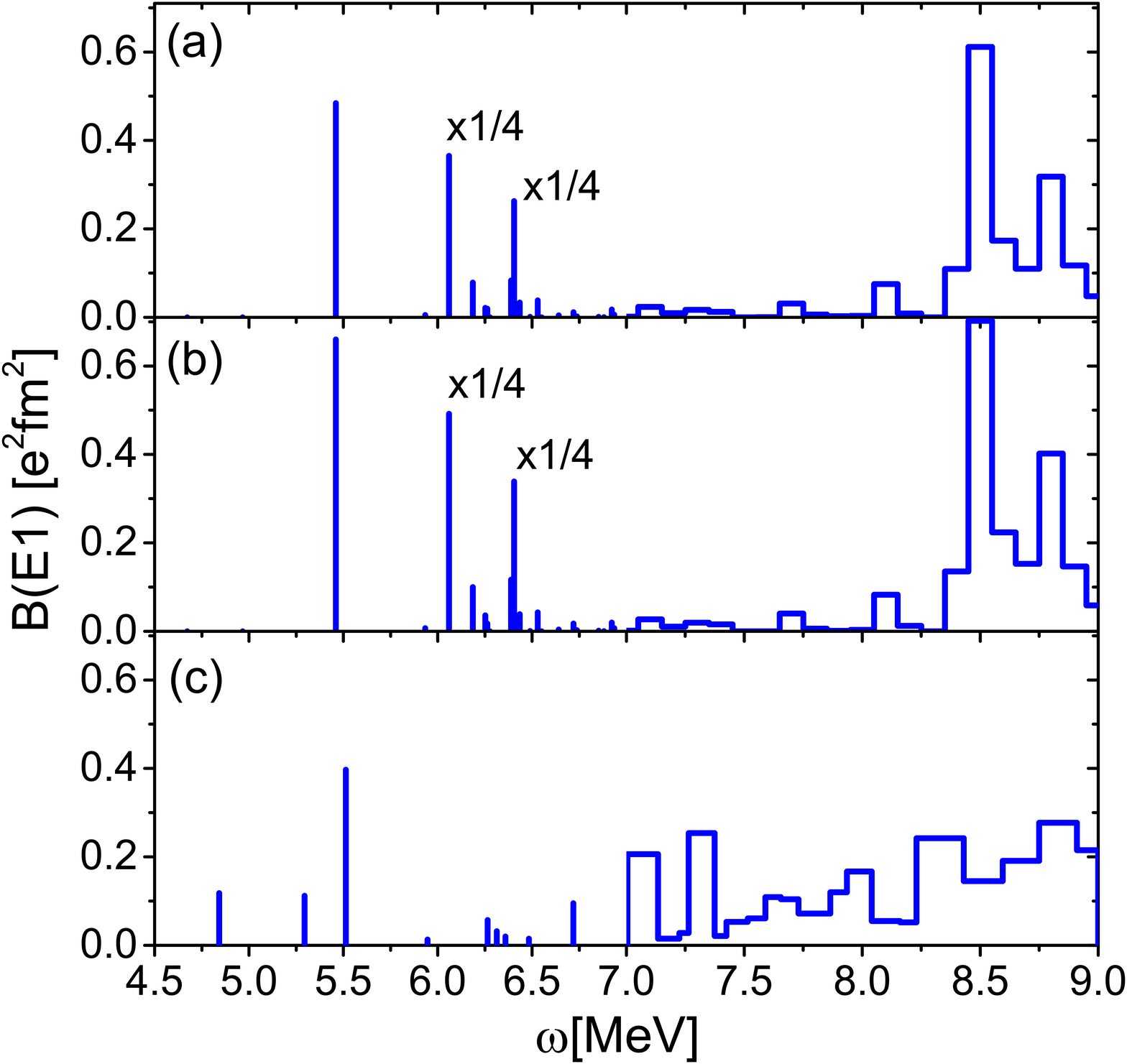}
\caption{ (Color online)  \label{fig6}  EMPM versus experimental $E1$ low-lying spectra. The EMPM spectra are compute with (a) and without (b) ground state correlations. The experimental data (c) are taken from
\cite{Polto}.
} 
\end{figure}

Both correlated and HF  ground states are used, in alternative, to compute the EMPM cross section. The two theoretical quantities  are compared with the experimental data
in Figure \ref{fig3}.   
With respect to TDA, shown in Figure \ref{fig1}, the two EMPM cross sections are  quenched   and follow  roughly the smooth trend
of the measured cross section.

The data are better reproduced if the unperturbed HF ground state $\mid 0 \rangle$ is used, while the correlations induce a too pronounced damping. In fact, in the HF case, the transition amplitude is simply given by the term  (\ref{M01ph}) which couples  $\mid 0 \rangle$ to the one-phonon components of the $1^-$ states. When the correlated ground state is used,   this term, though remaining dominant, is multiplied by the amplitude coefficient ${\cal C}^{(0^+)}_{0}$ of the HF component ( Eq.(\ref{M0l2})), which is of the order $|{\cal C}^{(0^+)}_{0}| \sim 0.8$. 

Such a quenching factor should be counterbalanced by a corresponding overall enhancement of the amplitudes of the one-phonon components. This, however, is   promoted  by the coupling to the three-phonon subspace, not included here. The conclusion to be drawn is that, in the absence of three phonons, the transition strengths should be computed by adopting the unperturbed HF ground state consistently with the assumption made for the energy levels. Both assumptions are tacitly made in all extensions of RPA. 

The calculation yields a secondary peak around $\sim 17$ MeV not observed experimentally. Such an unwanted  peak does not appear in other microscopic multiphonon calculations like the one carried out in QPM \cite{BertPon99} and may be an effect of the mentioned  discrepancies between HF and empirical single particle energies.

 The overall strength as well as the energy weighted sum $m_1$ remain practically unchanged
in going from TDA to the EMPM if the HF ground state is adopted. In both cases, the momentum $m_1$ overestimates the Thomas-Reiche-Kuhn (TRK) sum rule by a factor $\sim 1.7$. This factor would be reduced by washing out  the unwanted secondary high-energy peak. Once again, the HF energies come into play. 
 
Quenching and smoothness of the cross section are a consequence of the fragmentation of the strength which   enhances enormously  the density of levels and shortens the TDA peaks.  
As shown in Figure \ref{fig4}, pronounced transitions occur at low energy in both TDA and EMPM. They are responsible for the low-lying bump  in the cross section (Figure \ref{fig3}). The EMPM spectrum is extremely rich compared to TDA and is composed of shorter peaks (Figure \ref{fig5}). The quenching is considerably more pronounced when the ground state correlations are included. 

In both cases, 
the calculation reproduces only qualitatively the experimental strength distribution.  A detailed comparison (Figure \ref{fig6})
shows that significant discrepancies exist. Some EMPM transitions result to be either too strong or too weak compared to the measured ones.
The calculation, for instance, yields about 20 $1^-$ levels between $\sim 4.5$ MeV and $\sim 7$ MeV, twice as much as the levels detected experimentally \cite{Tamii11,Polto}.  The large majority of them, however, carry strengths which are either negligible  or of the order $10^{-2}$ $e^2 fm^2$. 

Most of the strength remains concentrated in three states. The transition at $\simeq 5.46$ MeV has a strength close to the one measured for the level at  $5.51$ MeV. The other two, at $\simeq 6.06$ MeV and $\simeq 6.40$ MeV,  do not have  experimental counterparts. As shown in Table \ref{tab1}, these three states have basically a one-phonon character.  Apparently they are little affected by the phonon coupling, which is, instead, very effective in the higher energy regions. 

In any case, one can infer from a comparison with Figure 6 of Ref. \cite{Polto} that our computed spectrum is  comparable with the ones computed in QPM and RTBA and, especially, with the spectrum obtained in shell model by  using empirical single particle energies and a phenomenological potential in a restricted configuration space \cite{Schweng10}.

 \begin{figure}[ht]
\includegraphics[width = \columnwidth]{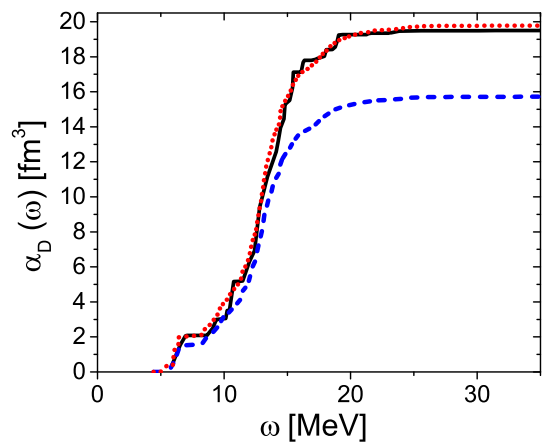}
\caption{ (Color online)  \label{fig7}  Polarizability as function of energy. The EMPM calculation is carried out using a HF (red line) and  a correlated ground state (black line).
} 
\end{figure}

It is of great interest to compute  the dipole polarizability  $\alpha_D$ (\ref{alfaD}).
If the HF ground state is used, we obtain $\alpha_D \simeq 19.78$ $fm^3$, very close to the value  $\alpha_D = 20.1 \pm 0.6$ fm$^3$ determined experimentally, once all the data up to 130 MeV are included \cite{Tamii11}.
Apparently, the phonon coupling does not affect the polarizability. This gets quenched, instead, if the ground state correlations are included 
and assumes the smaller value  $\alpha_D \simeq 15.71$ $fm^3$.
 
As shown in Figure \ref{fig7}, $\alpha_D(\omega)$ rises sharply in correspondence of the PDR and the GDR regions, indicating that it is determined  almost exclusively by these two resonances.

The dipole polarizability has been shown \cite{ReiNa10} to be closely correlated to  the excess neutron radius
\begin{equation} 
\Delta r_{np} = \sqrt{<r_n^2>} - \sqrt{<r_p^2>}.
\end{equation}
This was actually extracted from the measured $\alpha_D$ in $^{208}$Pb by exploiting such a correlation \cite{Tamii11}.   

We have computed the neutron skin radius using both HF and correlated ground states (\ref{Psi}). The EMPM yields for the proton ($\tau=p)$ and neutron ($\tau=n$) mean square radii the expression
\begin{eqnarray}
\label{rad1}
\langle r_\tau^2 \rangle = \langle \Psi_0 \mid  r^2_\tau \mid  \Psi_0 \rangle  = \langle r^2_\tau \rangle_{HF} + 
\langle r^2_\tau \rangle_{corr},
\end{eqnarray} 
where $\langle r^2_\tau \rangle_{HF}$ is the HF value and $\langle r^2_\tau \rangle_{corr}$ is the contribution coming from the two-phonon 
components. This is given by 
	\begin{eqnarray}
\label{rad2}
\langle r^2_\tau \rangle_{corr}   
= \sum_{\alpha \beta} \delta_{J_\alpha J_{\beta}} \delta_{J_\alpha 0}{\cal C}^{(0)}_{\alpha} {\cal C}^{(0)}_{\beta} 
 {\cal  M}_{\alpha \beta}^{(0)}, 
\end{eqnarray}
where
\begin{eqnarray}
\label{rad3}
{\cal  M}_{\alpha \beta}^{(0)}  = 
\sum_{r s} \langle r \parallel r^2_\tau  \parallel s \rangle    
\langle  \beta \parallel (a^\dagger_r \times  b_s)^0 \parallel  \alpha \rangle.
\end{eqnarray}
The HF proton radius is $r^{(HF)}_p = 5.06$ fm, smaller than the experimental value $r_p^{(exp)} \simeq 5.45$ fm. 
The neutron radius is estimated to be  $r^{(HF)}_n = 5.28$ fm. We, thus, obtain for the neutron skin $\Delta r_{np} = 0.22$ fm, which is in the range of the values determined experimentally \cite{Tamii11,Abrah12,Tarbert14}.
 
If the ground state correlations are taken into account, the proton radius raises to  $r_p = 5.19$ fm, not enough to reach the measured value. On the other hand, this estimate was computed in a truncated two-phonon space and, therefore, is to be considered a lower limit. An enlargement of the space would increment further the radius.   

The neutron radius raises from $r^{(HF)}_n = 5.28$ fm to   $r_n = 5.5$ fm. It follows that the neutron excess radius goes from the  $\Delta r^{(HF)}_{np} = 0.22$ fm to  $\Delta r_{np} = 0.31$ fm, which approaches the upper limit of the range of values deduced from experiments \cite{Abrah12}.  
 
\begin{figure}[ht]
\includegraphics[width = \columnwidth]{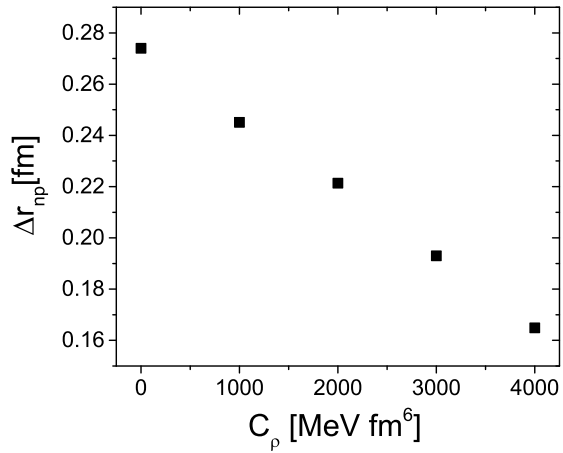}
\caption{  \label{fig8}  Neutron skin thickness versus the strength of $V_\rho$. 
} 
\end{figure}

Like  the density distribution (Figure \ref{fig2}), the radii are quite sensitive to the strength $C_\rho$ of the density dependent potential $V_\rho$. As shown in Figure \ref{fig8}, the  neutron excess radius decreases as $C_\rho$ increases. It would therefore be easy to reduce or enhance such a quantity by  a modest increment (decrement) of such a strength.

\begin{figure}[ht]
\includegraphics[width = \columnwidth]{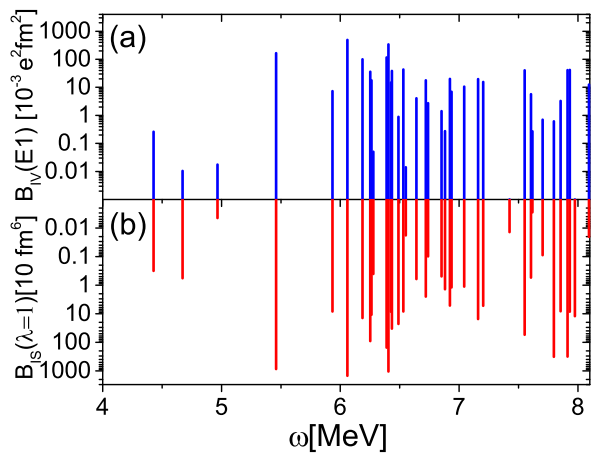}
\caption{ (Color online)  \label{fig9}  EMPM  isovector (a) versus isoscalar (b) dipole low-lying spectra
plotted on a logarithm scale so as to make clearly visible the many small peaks not discernible otherwise. 
The isovector strength is nothing but the low-lying sector of the spectrum shown in Figure \ref{fig4}(b).
The isoscalar one is obtained by using the isoscalar operator (\ref{E1IS}).
} 
\end{figure}

\begin{figure}[ht]
\includegraphics[width = \columnwidth]{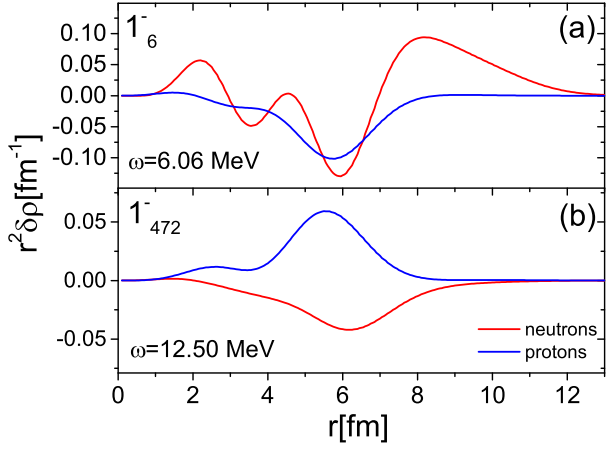}
\caption{ (Color online)  \label{fig10} 
$E1$ transition density for a low-lying   $1^-$ state (a) and one in the GDR region (b). }
\end{figure}

\begin{table}[th]
\caption{\label{tab1} Phonon composition of the lowest twenty $1^-$ states}
\begin{tabular}{ccccccc}
\hline \hline
$J^\pi_\nu$&&$\omega_\nu $ (MeV)&&    $|C^{(\nu)}_1|^2$ &&$|C^{(\nu)}_2|^2$  \\
\hline
 $1^-_1$  && 4.42780     && 0.00017    && 0.99983     \\  
 $1^-_2$  && 4.67271     && 0.00083    && 0.99917     \\ 
 $1^-_3$  && 4.96609     && 0.00014    && 0.99986     \\
 $1^-_4$  && 5.46012     && 0.95558    && 0.04442     \\
 $1^-_5$  && 5.93408     && 0.03132    && 0.96868     \\
 $1^-_6$  && 6.05979     && 0.90712    && 0.09288     \\
 $1^-_7$  && 6.18594     && 0.05422    && 0.94578     \\
 $1^-_8$  && 6.25179     && 0.04936    && 0.95064     \\
 $1^-_9$  && 6.26285     && 0.05409    && 0.94591     \\
 $1^-_{10}$ && 6.27701     && 0.00310    && 0.99690     \\
 $1^-_{11}$ && 6.38869     && 0.15931    && 0.84069     \\
 $1^-_{12}$ && 6.40474     && 0.69907    && 0.30093     \\
 $1^-_{13}$ && 6.42531     && 0.03371    && 0.96629     \\
 $1^-_{14}$ && 6.43502     && 0.03215    && 0.96785     \\
 $1^-_{15}$ && 6.48971     && 0.86985    && 0.13015     \\
 $1^-_{16}$ && 6.53002     && 0.00956    && 0.99044     \\
 $1^-_{17}$ && 6.55127     && 0.00485    && 0.99515     \\
 $1^-_{18}$ && 6.64103     && 0.00346    && 0.99654     \\
 $1^-_{19}$ && 6.71925     && 0.01301    && 0.98699     \\
 $1^-_{20}$ && 6.73778     && 0.00058    && 0.99942     \\ 
 \hline \hline  
 \end{tabular}
\end{table}

\begin{table}[th]
\caption{\label{tab2} Particle-hole composition of two selected phonons}
\begin{tabular}{cccc}
$1^-$& $\omega =  5.890 MeV$ & &  \\

\hline \hline
$(p(h)^{-1})^\pi$ & $C_{ph}^\pi$ & $(p(h)^{-1})^\nu$   & $C_{ph}^\nu$   \\

\hline

 $0h_{9/2}      (0g_{7/2})^{-1}$      &  -0.05221  &          $0i_{11/2}     (0h_{9/2})^{-1}  $  &   0.05390  \\
 $1f_{7/2}      (1d_{5/2})^{-1}$      &  -0.03103  &          $1g_{9/2}      (1f_{7/2})^{-1}  $  &  -0.05141  \\
 $1f_{5/2}      (1d_{3/2})^{-1}$      &  -0.02924  &          $2d_{5/2}      (1f_{7/2})^{-1}  $  &   0.01384  \\
 $2p_{3/2}      (1d_{5/2})^{-1}$      &   0.01366  &	        $2d_{5/2}      (2p_{3/2})^{-1}  $  &  -0.04462  \\
 $2p_{1/2}      (1d_{3/2})^{-1}$      &   0.01046  &          $3s_{1/2}      (2p_{3/2})^{-1}  $  &   0.15998  \\
 $2p_{1/2}      (2s_{1/2})^{-1}$      &  -0.01926  &          $3s_{1/2}      (2p_{1/2})^{-1}  $  &   0.97117  \\
 $0i_{13/2}     (0h_{11/2})^{-1}$     &  -0.06687  &          $2d_{3/2}      (2p_{3/2})^{-1}  $  &  -0.02626  \\
																			&            &          $2d_{3/2}      (2p_{1/2})^{-1}  $  &   0.06317  \\
																			&            &          $1g_{7/2}      (1f_{5/2})^{-1}  $  &  -0.03117  \\																			&            &          $0j_{15/2}     (0i_{13/2})^{-1} $  &   0.08284  \\																			&            &          $2g_{9/2}      (1f_{7/2})^{-1}  $  &   0.01165  \\																			&            &          $3d_{5/2}      (2p_{3/2})^{-1}  $  &   0.01132  \\																			&            &          $4s_{1/2}      (2p_{3/2})^{-1}  $  &  -0.01727  \\
																			&            &          $2g_{7/2}      (1f_{5/2})^{-1}  $  &   0.01277  \\																																						

\hline
$1^-$& $\omega =  12.816 MeV$ & &  \\
\hline

 $0h_{9/2}    (0g_{9/2})^{-1}$   & -0.11797    &        $0i_{11/2}     (0h_{9/2})^{-1}$    &  -0.12090  \\
 $0h_{9/2}    (0g_{7/2})^{-1}$   & -0.09381    &        $1g_{9/2}      (0h_{9/2})^{-1}$    &   0.08459  \\
 $1f_{7/2}    (0g_{7/2})^{-1}$   & -0.03875    &        $1g_{9/2}      (1f_{7/2})^{-1}$    &  -0.21304  \\
 $1f_{7/2}    (1d_{5/2})^{-1}$   &  0.07336    &        $2d_{5/2}      (1f_{7/2})^{-1}$    &   0.01687  \\
 $1f_{5/2}    (0g_{7/2})^{-1}$   & -0.03795    &        $2d_{5/2}      (1f_{5/2})^{-1}$    &  -0.02538  \\
 $1f_{5/2}    (1d_{5/2})^{-1}$   & -0.09910    &        $2d_{5/2}      (2p_{3/2})^{-1}$    &  -0.03684  \\
 $1f_{5/2}    (1d_{3/2})^{-1}$   &  0.04318    &        $3s_{1/2}      (2p_{1/2})^{-1}$    &  -0.01597  \\
 $2p_{3/2}    (1d_{3/2})^{-1}$   &  0.17526    &        $2d_{3/2}      (1f_{5/2})^{-1}$    &  -0.03881  \\
 $2p_{3/2}    (2s_{1/2})^{-1}$   &  0.02467    &        $2d_{3/2}      (2p_{3/2})^{-1}$    &  -0.03722  \\
 $2p_{1/2}    (1d_{3/2})^{-1}$   & -0.17532    &        $2d_{3/2}      (2p_{1/2})^{-1}$    &   0.01313  \\
 $0i_{13/2}   (0h_{11/2})^{-1}$  &  0.46993    &        $1g_{7/2}      (0h_{9/2})^{-1}$    &   0.04067  \\
 $1g_{9/2}    (0h_{11/2})^{-1}$  & -0.01005    &        $1g_{7/2}      (1f_{7/2})^{-1}$    &  -0.57127  \\
 $0i_{11/2}   (0h_{11/2})^{-1}$  & -0.06708    &        $1g_{7/2}      (1f_{5/2})^{-1}$    &   0.03749  \\
 $3p_{3/2}    (1d_{3/2})^{-1}$   &  0.01351    &        $0j_{15/2}     (0i_{13/2})^{-1}$   &   0.45810  \\
 $3p_{3/2}    (2s_{1/2})^{-1}$   &  0.01245    &        $1h_{11/2}     (0i_{13/2})^{-1}$   &  -0.01488  \\
 $                           $   &             &        $2g_{9/2}      (1f_{7/2})^{-1}$    &  -0.05808  \\
 $                           $   &             &        $3d_{5/2}      (1f_{5/2})^{-1}$    &  -0.03141  \\
 $                           $   &             &        $3d_{5/2}      (2p_{3/2})^{-1}$    &  -0.08047  \\
 $                           $   &             &        $4s_{1/2}      (2p_{1/2})^{-1}$    &  -0.06666  \\
 $                           $   &             &        $3d_{3/2}      (1f_{5/2})^{-1}$    &  -0.04485  \\
 $                           $   &             &        $3d_{3/2}      (2p_{3/2})^{-1}$    &  -0.05249  \\
 $                           $   &             &        $3d_{3/2}      (2p_{1/2})^{-1}$    &   0.02462  \\
 $                           $   &             &        $2g_{7/2}      (0h_{9/2})^{-1}$    &  -0.01030  \\
 $                           $   &             &        $2g_{7/2}      (1f_{7/2})^{-1}$    &  -0.05356  \\
 $                           $   &             &        $0j_{13/2}     (0i_{13/2})^{-1}$   &  -0.18118  \\

\hline
\hline
 \end{tabular}
\end{table}

For a more complete investigation of the nature of the low-lying $E1$ levels, we compute also the isoscalar dipole transition strength using the operator 
\begin{eqnarray} 
{\cal M}_{IS} (\lambda =1, \mu) = 
\sum_{i=1}^A   r_i^{3} Y_{1 \mu}(\hat{r_i}). 
\label{E1IS}
\end{eqnarray}
The computed isoscalar and  isovector spectra overlap over the entire low-energy region (Figure \ref{fig9}), in perfect analogy with the equivalent calculation performed for $^{132}$Sn \cite{knap14}.

This prediction seems to be disproved by the  isoscalar-isovector splitting observed
 in $(\alpha, \alpha' \gamma)$ experiments on open shell nuclei like $^{140}$Ce \cite{Savran06} and $^{124}$Sn
 \cite{Gov98,Endres10,Endres12}.
For a conclusive test, it would be of great help to perform an analogous experiment on the stable doubly magic  $^{208}$Pb.

Our theoretical results suggests that these low-lying transitions are promoted mainly by the excitations of the neutrons above the N=Z core thereby hinting at  their pygmy nature.

Such a hint is confirmed by the behavior of the transition densities,
shown in Figure \ref{fig10}. The low-lying transition  density shows a neutron excess for large values of the radial coordinate. This behavior is to be contrasted with the one exhibited by  the transition to the most  strongly excited $1^-$ state in the region of the GDR, which clearly describes an oscillation of protons versus neutrons. 
 
A further support comes from the investigation of the structure of the wave functions.
As shown in Table \ref{tab1}, the most strongly excited low-lying states have an overwhelming one-phonon component. The large majority of the  weakly excited states, instead, have a dominant two-phonon structure.    
 
Table \ref{tab2} shows  the p-h composition of two typical phonons  contributing to the strength in the low-energy  and GDR regions, respectively. They have  distinctive   structures.  The low-lying phonon is composed predominantly of p-h neutron excitations above the $N=Z$ core, with a very dominant single p-h neutron configuration. The  phonon   in the GDR region   is built of   both proton and neutron p-h configurations of comparable amplitudes in opposition of phase.

\section{Concluding remarks}
The inclusion of a subset of the two-phonon basis states in the selfconsistent EMPM calculation  fragments strongly  the dipole strength computed in TDA by greatly enhancing the density  of levels. 

It is, however, necessary to assume an unperturbed HF ground state  in order to reproduce with fairly good accuracy the GDR in its magnitude and smooth trend and to describe satisfactorily the global properties of the dipole response, described here by the dipole polarizability and the neutron skin radius.

This assumption, which underlies all extensions of RPA, has a theoretical justification. Since the HF ground state couples strongly to 
the two-phonon basis,   it is necessary to include the three-phonon states in order to obtain a comparable effect on the $1^-$ one-phonon  states. Consistency requires that  the two-phonon correlations in the ground state should be neglected if the three-phonon basis is missing.

The ultimate goal is to enlarge the space so as to include such a basis. 
The coupling of the three-phonons  to the other subspaces is expected to  modify the weight of the one and two phonon components of the total $1^-$ wave functions and, therefore, should affect the strength distribution. Whether such a redistribution leads to a better agreement with experiments can be ascertained only by explicit calculations.  
In this perspective, we are trying to improve the efficiency of the codes and, concurrently, search for reliable methods for truncating  the phonon space.

It is also compelling  to improve further the agreement between HF and empirical single particle spectra in order to be able to get an unambiguous correspondence between theoretical and experimental levels, especially in the low-energy region. 
Such a task can be achieved only by  acting on the  NN interaction.

Though the optimized chiral potential  $V_\chi = NNLO_{opt}$\cite{Ekstr13} may represent a promising  starting point,
higher order terms need to be taken into account. This is done here effectively by adding a phenomenological density dependent potential which simulates a contact three-body force.

Such a potential plays a crucial role. It improves to some extent the description of the single particle levels,  enhances the diffuseness of the Fermi surface consistently with experiments, and shifts the dipole spectrum in the region of observation. It is, however, phenomenological and contains an unconstrained coupling constant.

We need, in any case, a more refined potential able to provide a more accurate and detailed description of the single particle spectra.  The new optimized interaction $NNLO_{sat}$ \cite{Ekstr15}, involving  both two- and three-body components of NNLO, is a possible candidate.

A calculation based on  such a potential would be parameter free and  would link the dipole spectra directly and exclusively  to the   NN interaction.  If carried out in a space encompassing up to three phonons,  it would represent an additional reliable test for  the chiral interaction.

\begin{acknowledgments}
This work was partly supported by the Czech Science Foundation (project no. P203-13-07117S).
Two of the authors (F. Knapp and P. Vesel\'{y}) thank the Istituto Nazionale di Fisica Nucleare (INFN) for  financial support.  
Highly appreciated was the access to computing and storage facilities 
provided by the MetaCentrum under the program LM2010005 and the CERIT-SC under the program Centre CERIT Scientific Cloud, part of the Operational Program Research and Development for Innovations, Reg. no. CZ.1.05/3.2.00/08.0144.
\end{acknowledgments}

\bibliographystyle{apsrev}
\bibliography{Pb208EMPMSelf}
\end{document}